\documentclass[a4paper,10pt]{article}
\usepackage[english]{babel}
\usepackage[pdftex]{graphicx}

\usepackage{hyperref}

\begin{document}


\title{Modeling the emergence of a new language:\\
Naming Game with hybridization\footnote{The final publication will be available at
\url{http://www.springer.com/lncs}}}
\author{Lorenzo~Pucci$^1$, Pietro~Gravino$^{2,3}$, Vito~D.P.~Servedio$^2$\\[3mm]
{\small $^1$Phys.~Dept., Univ.~Federico II, Complesso Monte~S.~Angelo, 80126 Napoli, Italy}\\
{\small $^2$Phys. Dept., Sapienza Univ. of Rome, P.le A. Moro 2, 00185 Roma, Italy}\\
{\small $^3$Phys. Dept., Alma Mater Studiorum, Univ. of Bologna, Italy}}

\maketitle

\begin{abstract}
{\noindent In recent times, the research field of language dynamics has focused on the
investigation of language evolution, dividing the work in three evolutive steps,
according to the level of complexity: lexicon, categories and grammar.
The Naming Game is a simple model capable of accounting for the emergence of a
lexicon, intended as the set of words through which objects are named. We
introduce a stochastic modification of the Naming Game model with the aim of
characterizing the emergence of a new language as the result of the interaction
of agents. We fix the initial phase by
splitting the population in two sets speaking either language A or B. Whenever
the
result of the interaction of two individuals results in an agent
able to speak both A and B, we introduce a finite probability that this state
turns into a new idiom C, so to mimic a sort of hybridization process. 
We study the system in the space of parameters
defining the interaction, and show that the proposed model displays
a rich variety of behaviours, despite the simple mean field topology
of interactions.
}
\end{abstract}

\section{Emergence of a lexicon as a language}

The modeling activity of language dynamics aims at describing language evolution
as the global effect of the local interactions between individuals in a
population of $N$ agents, who tend to align their verbal behavior locally, by a
negotiation process through which a successful communication is achieved \cite{loretoreview,Loreto2011}. 
In this framework, the emergence of a
particular communication system is not due to an external coordination, or a
common psychological background, but it simply occurs as a convergence effect in
the dynamical processes that start from an initial condition with no existing
words (agents having to invent them), or with no agreement.

\begin{figure}[!t]
 \centering
\includegraphics[width=10cm]{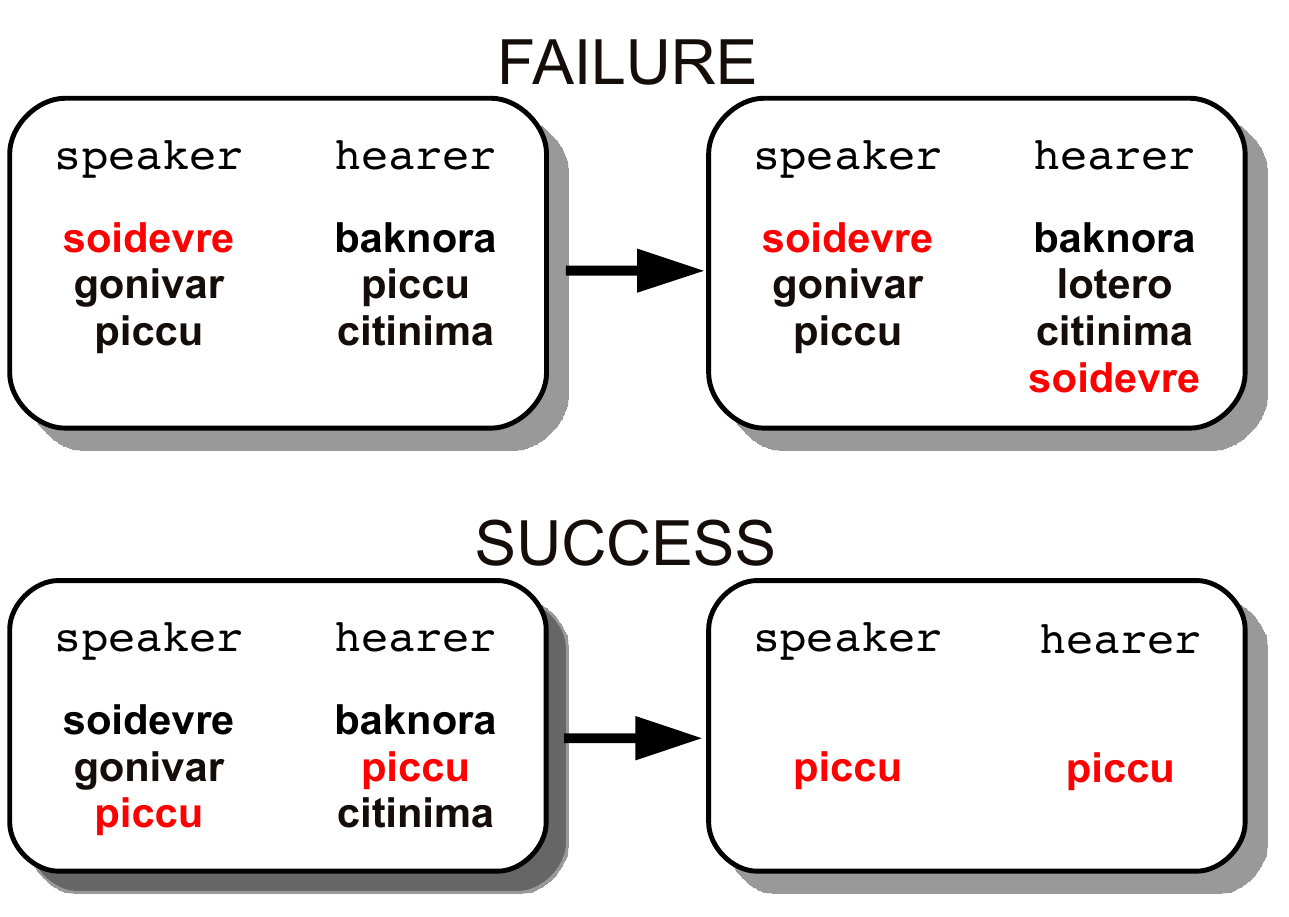}
\caption{\textbf{Naming Game interaction scheme.} A \emph{speaker} and a
\emph{hearer} are
picked up randomly. The speaker utters a word chosen randomly in his vocabulary.
\emph{Failure}: the hearer does not know the uttered word and he simply adds it
to his vocabulary. \emph{Success}: the hearer knows the uttered word and both
agents by agreement delete all the words in their vocabularies except the
uttered word one.}
\label{int-base}
\end{figure}

Our work is based on the Naming Game (NG) model, and on its assumptions~\cite{Steels1995}. 
In Fig.~\ref{int-base} we recall the NG basic pairwise interaction scheme.
A fundamental assumption of NG is that vocabulary evolution associated to every
single object is considered independent. This lets us simplify the evolution of
the whole lexicon as the evolution of the set of words associated to a single
object, equally perceived in the sensorial sphere by all
agents.

The simplicity of NG in describing the emergence of a lexicon also
relies on the fact that competing words in an individual vocabulary
are not weighted, so that they can be easily stored or deleted~\cite{Baronka2006}. 
In fact it turns
out that for convergence to a single word consensus state, the weights are not
necessary as it was supposed by the seminal work in this
research field~\cite{Steels1995}. 
Every agent is a box that could potentially contain an infinite
number of words, so the number of states that can define the agent is limited
only by the number of words diffused in the population at the beginning of the
process (so that anyone can speak with at least one word).

In this work we aimed not only at the aspect of competition of a language form
with other ones, but also at introducing interactions between them, with the possibility of
producing new ones. We investigate conditions for the success of a new idiom, as
product of a synthesis and competition of preexisting idioms. 
To this purpose, we introduce a stochastic
interaction scheme in the basic Naming Game accounting for this synthesis, and
show in a very simple case that the success of the new spoken form at expense of
the old ones depends both on the stochastic parameters and the fractions of
the different idioms spoken by populations at the beginning of the process. We
have simulated this process starting from an initial condition where a 
fraction $n_A$ of
the population speaks with $A$  and the remaining $n_B$ with $B$.
It turns out that when the different-speaking fractions are roughly of the same size the
new form, which we call $C$ (therefore we shall refer to our model as the 
``ABC model'' in the following), created from the synthesis of $A$ and $B$, establishes and
supplants the other two forms. Instead, when $n_A>n_B$ (or symmetrically
$n_B>n_A$), above a threshold depending on the chosen stochastic parameters, the term $A$
establishes (or symmetrically $B$), namely one of the starting idioms prevails
and settles in the population.

Previous attempts to model the emergence of a Creole language, i.e.\ an idiom originating by a sort of hybridization and fusion of languages, can be found in literature \cite{satterfield,nakamura,Jansson}.

\section{The ABC Model}

\begin{figure}[!t]
\centering
\includegraphics[width=10.5cm]{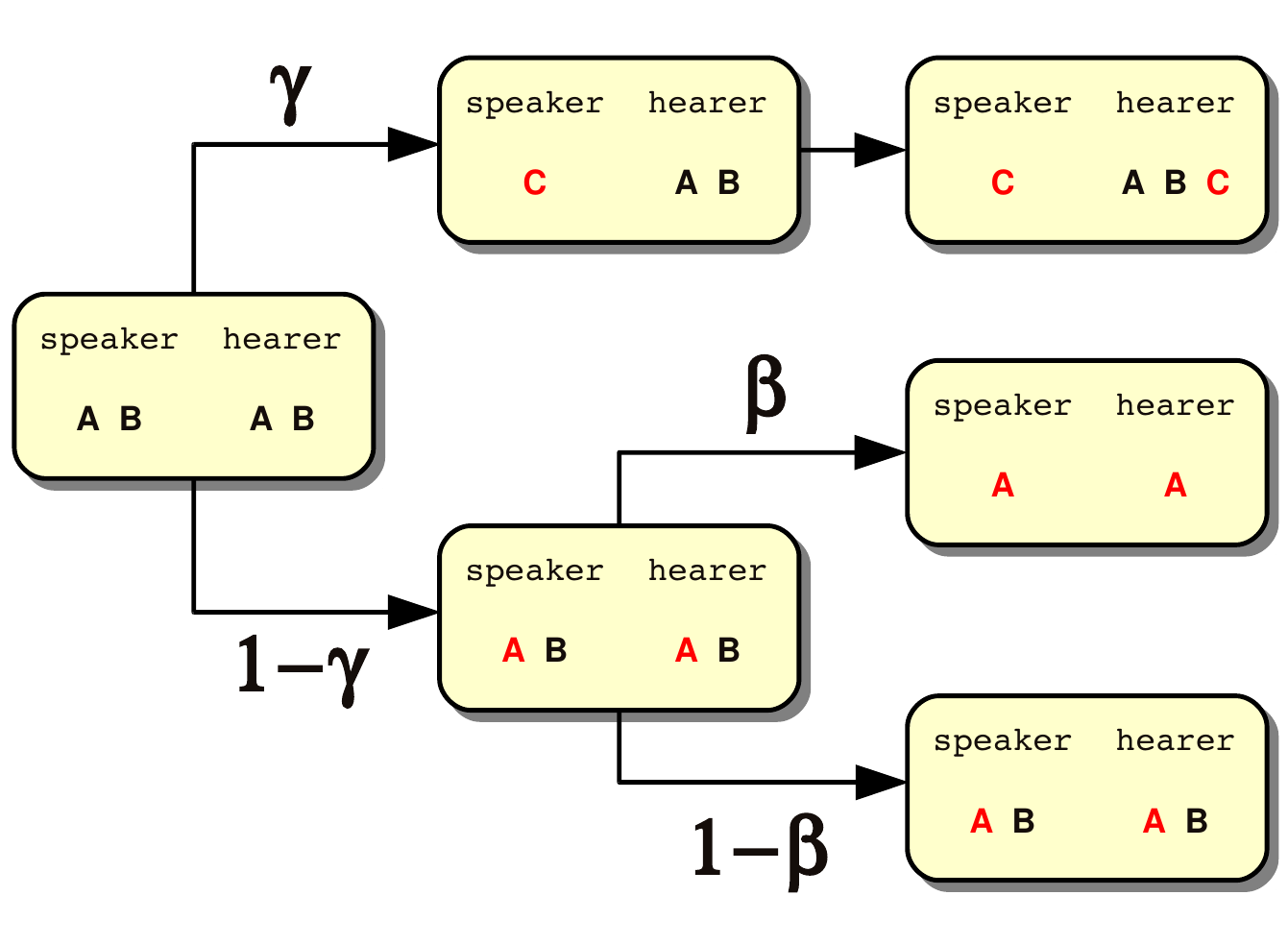}
\caption{\textbf{ABC model:}
Example of interaction with the introduction of the trasformation of $AB
\rightarrow C$ with probability $\gamma$ on the speaker vocabulary. The
trasformation is carried before the usual speaker-hearer NG interaction, which
includes the stochastic update of the vocabularies in case of a success
depending on $\beta$.}
\label{es-inter}
\end{figure}

The model we propose here is based on a mean field topology involving
$N$ agents, i.e.\ any two agents picked up randomly can interact. 
Despite this pretty simple topology of
interactions, the proposed model will show a richness of behaviors.
In the basic Naming Game, the initial condition is fixed with an empty word list
for each agent~\cite{Baronka2006}. If an agent chosen as speaker still owns an
empty list, he invents a new word and utters it. 
In our proposed new model all
agents are initially assigned a given word, either $A$ or $B$, so that there is no
possibility to invent a brand new word, unless an agent with 
both $A$ and $B$ in his list is selected. In that case,
we introduce a finite probability $\gamma$ that his list containing $A$ and $B$
turns into a new entry $C$ (Fig.~\ref{es-inter}). We interpret $\gamma$ as a
measure of the need of agents to obtain a sort of hybridization through which 
they understand each other, or a measure of the natural tendency of two
different language forms to be synthesized together. 
In the latter case, different
values of $\gamma$ would depend on the language forms considered in a real
multi language scenario.

The stochastic rule $AB\rightarrow C$ may be applied either in the initial interaction phase by
changing the dictionary of the speaker, or at the final stage by changing the
state of the hearer after the interaction with the speaker. In this paper we
show the results obtained by running the trasformation $AB\rightarrow C$ (and
also $ABC \rightarrow C$ when a speaker has got an $ABC$ vocabulary) before the
interaction. Introducing the trasformation after the interaction changes
the results from a qualitative point of view, producing only a shift of
transition lines between the final states in the space of parameters defining the
stochastic process.

A further stochastic modification of the basic Naming Game interaction, firstly
introduced in~\cite{BaronchelliBeta}, has also been adopted here. It gives account for
the emergence or persistance of a multilingual final state, where more than one
word is associated to a single object. This is done by mimicking a sort of
confidence degree among agents: in case of a successful interaction, namely when
the hearer shares the word uttered by the speaker, the update trasformation of
the two involved vocabularies takes place with probability $\beta$ (the case
$\beta=1$ obviously reduces to the basic NG).
Baronchelli et al.~\cite{BaronchelliBeta} showed in their model 
(which corresponds to our model at $\gamma=0$) that a transition occurs around $\beta=1/3$.
For $\beta>1/3$ the system converges to the usual one word consensus state (in
this case only $A$ or $B$). 
For $\beta<1/3$ the system converges to a mixed
state, where more than one word remains, so that there exist single and multi
word vocabularies at the end of the process (namely $A$, $B$ and $AB$). 
A linear stability analysis of the mean field master equations of this model
(describing the evolution of vocabulary frequencies $n_A$, $n_B$ and $n_{AB}$)
shows that the steady state $n_A=1$, $n_B=n_{AB}=0$ (or symmetrically $n_B=1$,
$n_A=n_{AB}=0$), which is stable for $\beta>1/3$, turns unstable if
$\beta<1/3$, where viceversa the steady state
$n_A=n_B=b\left(\beta\right)$, $n_{AB}=1-2b\left(\beta\right)$ emerges as a stable
state,
with $b(\beta)$ being a simple algebraic expression of the parameter $\beta$.
In our work, as shown next, we found that the transition order-disorder 
(i.e.\ single word vs.\ multi word final state) at $\beta=1/3$ remains for all the
values of $\gamma$. 

\begin{figure}[!t]
\centering
\includegraphics[width=60mm]{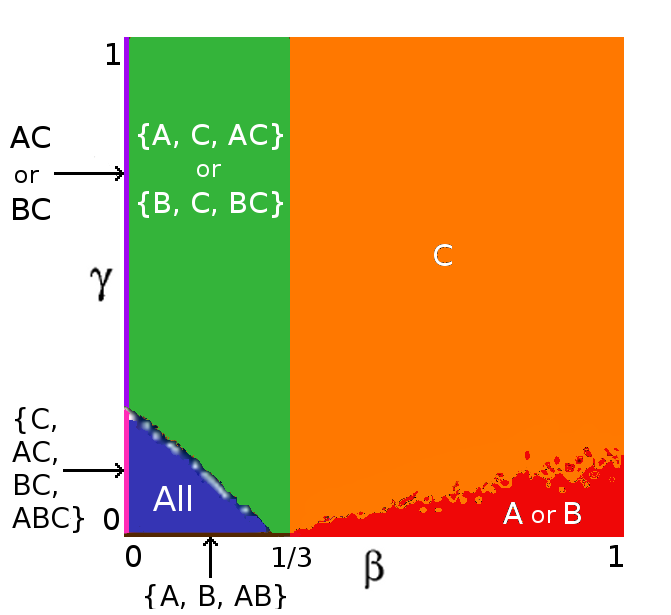}%
\includegraphics[width=60mm]{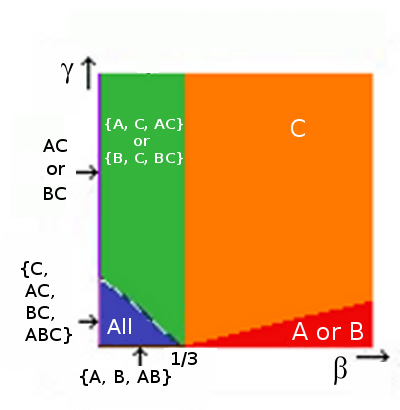}
\caption{\textbf{\boldmath Phase diagram $\gamma,\beta$ of the ABC model:} 
The new language $C$ is created at the beginning
of the interaction by updating the speaker's vocabulary. 
We have $0\le\beta\le1$ and
$0\le\gamma\le1$ on the horizontal and vertical axis, respectively. 
\emph{Left:} 
Results of the stochastic model where the number
of agents involved is $N=1000$ and the final time is set to $10^{6}$. The
initial condition on populations is fixed as $n_A\simeq n_B$ and $n_{AB}=0$ at
$t=0$.
\emph{Right:} 
Results of the mean field master equation numerically integrated till a final time $t=1000$. 
The initial condition where chosen as $n_A=0.49$ and $n_B=0.51$ (and symmetrically
$n_A=0.51$ and $n_B=0.49$). 
We employed a fourth order Runge-Kutta algorithm with
temporal step $dt=0.1$. }
\label{df-10000}
\end{figure}

The numerical stochastic simulation of the process for selected values 
of $\left( \beta,\gamma\right)\in\left[0,1\right]\times\left[0,1 \right]$, 
indicates that the system presents a varied final phase
space as shown in Fig.~\ref{df-10000} and \ref{g-al-b07}, left panel. The transition line at
$\beta=1/3$ remains: for $\beta>1/3$ the system converges to a
one-word final state, with a single word among $A$, $B$ and $C$, while for
$\beta<1/3$ it converges to a multi-word state with one or more than one word spoken
by each agent.
This result is confirmed by the integration of the mean field master equation of the model,
describing the temporal evolution of the fractions of all vocabulary species
$n_A$, $n_B$, $n_C$, $n_{AB}$, $n_{AC}$, $n_{BC}$, $n_{ABC}$, present in the system. 
The results of such integration, involving a fourth order Runge-Kutta algorithm,
display the same convergence behaviour of the stochastic model 
(Fig.~\ref{df-10000} and \ref{g-al-b07}, right panel), though they are 
obviously characterized by less noise.

\subsection{High confidence \texorpdfstring{$\beta > 1/3$}{beta > 1/3}}

\begin{figure}[!t]
\centering
\includegraphics[width=6cm]{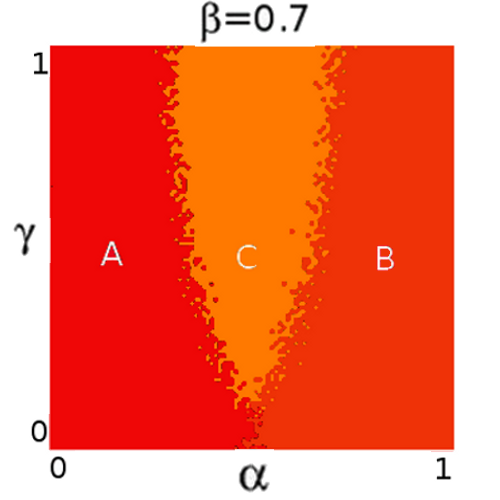}%
\includegraphics[width=6cm]{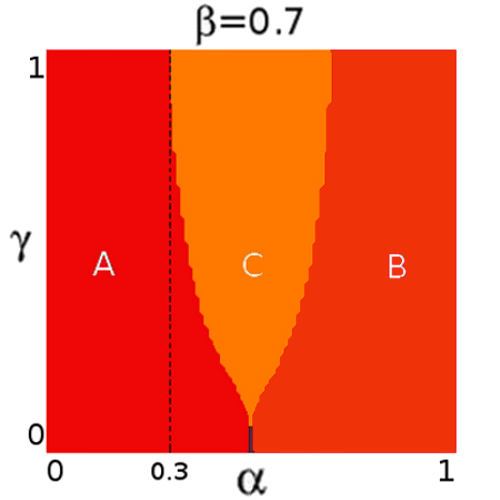}
\caption{\textbf{\boldmath Phase diagram $\gamma,\alpha$ of the ABC model:}
Final phase diagram depending on $\gamma$ and initial condition
$n_A=1-\alpha$ and $n_B=\alpha$ with $\alpha\in\left[ 0,1\right] $, with
$\beta=0.7$ fixed. 
\emph{Left:} diagram obtained after $10^6$ pair interactions of the stochastic process. 
\emph{Right:} diagram obtained by integrating up to $t=1000$ the mean field 
master equation
with a fourth order Runge-Kutta algorithm employing a temporal step $dt=0.1$.
For $\alpha<0.3$ (or symmetrically $\alpha>0.7$) the system
converges to the $n_A=1$ (or symmetrically $n_B=1$) for every considered value of
$\gamma$.}
\label{g-al-b07}
\end{figure}

By looking at Fig.~\ref{df-10000} at the region with $\beta>1/3$,
we note a transition interval between the final state composed 
only of either $A$ or $B$ (red region) and a final state with only $C$ (orange region). 
The
fuzziness of the border dividing these two domains, evident in the left panel of the figure,
can be ascribed to finite size effects, 
for the separation line gets sharper by enlarging the number of agents $N$, 
eventually collapsing towards the strict line obtained by the integration
of the mean field master equation (right panel of the figure), which we report in
the Appendix section. 
The linear stability analysis of the cumbersome mean field master equation reveals
that these two phases are both locally stable, and turn unstable when $\beta<1/3$.
The convergence to one or the other phase depends on the initial conditions, i.e.\
whether the system enters the respective actraction basins during its 
dynamical evolution.
To demonstrate this, we studied the behaviour of the system 
by fixing $\beta$ and varying both $\gamma$ and
the initial conditions $n_A=\alpha$, $n_B=1-\alpha$, with $\alpha\in\left[0,1\right]$.
The result reported in Fig.~\ref{g-al-b07} clearly shows a dependence on the
initial conditions. In particular, if the initial condition
$|n_A-n_B|=1-2\alpha$ is sufficiently large, the convergence to the $C$ phase
disappears.
The corresponding threshold value of $\alpha$ decreases slightly by decreasing
the value of parameter $\beta$, going from $\alpha=0.34$ for $\beta=1$ to
$\alpha=0.24$ for $\beta=0.34$ (i.e.\ slightly above the transition 
signalled by $\beta=1/3$).

\begin{figure}[!t]
\centering
\includegraphics[width=12cm]{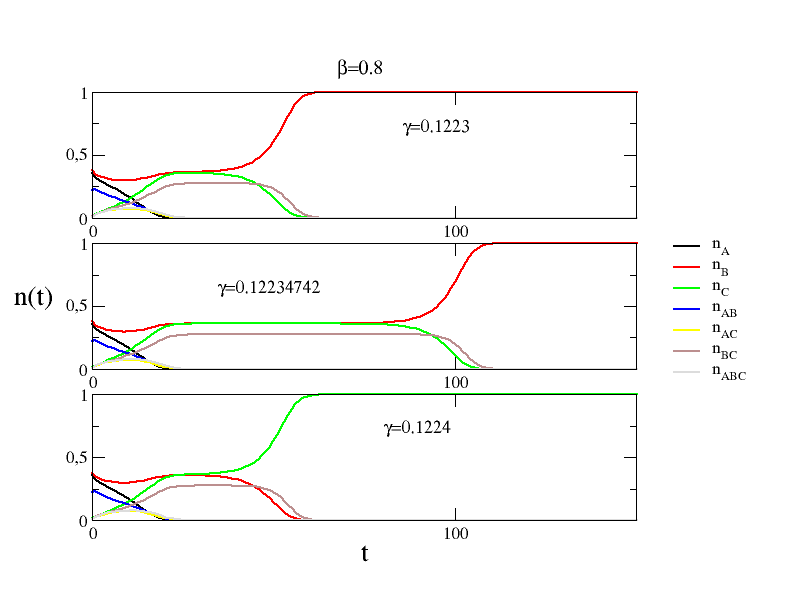}
\caption{\textbf{Temporal evolution of the ABC model:}
Temporal evolution of all vocabulary species occurring in the system
calculated by integrating the mean field master equation.
The initial population was chosen as $n_B=0.51$ and $n_A=0.49$, while we fixed
$\beta=0.8$. 
For $\gamma=0.1223$ the system converges to the $B$ consensus state, 
while for $\gamma=0.1224$ the system converges to the $C$ consensus 
state. 
The initial trivial transient with the formation of the $AB$ mixed states has been removed.
By approaching the transition point the convergence time increases.}
\label{n_t}
\end{figure}

By numerically solving the mean field master equation, we analyzed the 
evolution of the system in proximity of the transition between the (orange)
region characterized by the convergence to the $C$ state 
and the (red) region where the convergence is towards either the state $A$
 or $B$, being these latter states discriminated by the initial 
conditions $n_A>1/2$ or $n_A<1/2$ respectively.
We show the result in Fig.~\ref{n_t} obtained by fixing $\beta=0.8$ and $\alpha=0.49$.
Initially, both the fractions of $n_A$ (black curve) and $n_B$ (red curve) decrease 
in favour of $n_{AB}$ (blue curve).
Thereafter, also $n_{AB}$ starts to decrease since the mixed $AB$ and $ABC$
states can be turned directly into $C$, causing an increase of $n_C$ (green curve).
While the $n_{AB}$, $n_{ABC}$, $n_A$, $n_{AC}$ fractions vanish quite soon, 
mainly because fewer agents have the $A$ state in their vocabulary, 
the states involving $B$ and $C$ survive, reaching a meta-stable situation in which
$n_C=n_B\approx 0.37$ and $n_{BC}\approx 0.26$.
This meta-stable state of the system is clearly visible in the mid panel
of Fig.~\ref{n_t}.
The life time of the meta-stable state diverges by approaching the corresponding
set of parameters ($\gamma=\gamma_c\approx 0.12234742$ in the Figure; $\gamma_c$ depends on
$\beta$), 
with the result that the overall convergence time
diverges as well.
The stochastic simulation would differ from the solution of the master equation
right at this point: a small random fluctuation in the fraction of either
$B$ or $C$ will cause the convergence of the stochastic model towards one or the other state,
while the deterministic integration algorithm (disregarding computer numerical errors)
will result always in the win of the state
$C$ for $\gamma>\gamma_c$ or the state $B$ for $\gamma<\gamma_c$.
The fuzziness visible in all the figures related to the stochastic model
is a direct consequence of those random fluctuations.

\begin{figure}[!t]
\centering
\includegraphics[width=9cm]{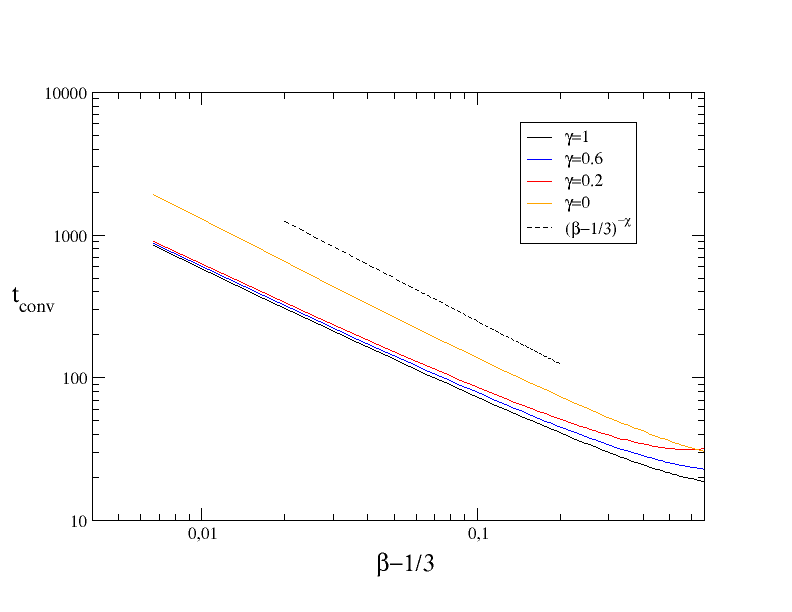}
\caption{
\textbf{\boldmath Convergence time in the ABC model for $\beta\rightarrow 1/3^+$:}
Time of convergence of the system to the one word consensus state
as obtained with the mean field master equation as a function of $(\beta-{1}/{3})$. 
The initial population densities were set to $n_A=0.49$ and $n_B=0.51$.
Importantly, if 
$\gamma>0$ the system converges to the hybrid state $C$, while for $\gamma=0$
the state $B$ prevails (orange curve). The dotted line refers to a power-law behaviour
of exponent $-1$. 
We used a fourth order Runge-Kutta algorithm with temporal step $dt=0.1$. }
\label{t_conv}
\end{figure}

Another interesting area in the $\gamma,\beta$ phase space of the model is the boundary
between the region around $\beta=1/3$, where one switches from a convergence to a single
state ($\beta>1/3$) to a situation with the coexistence of multiple states ($\beta<1/3$).
As $\beta\rightarrow {1}/{3}^+$ the time of
convergence towards the $C$ consensus phase, which is the
absorbing state whenever $\gamma>0$, diverges following a power-law with the
same exponent as in the case of $\gamma=0$, where we recover the results of 
\cite{BaronchelliBeta}, i.e.\ $t_{\mathrm{conv}}\simeq (\beta-1/3)^{-1}$ 
(Fig.~\ref{t_conv}).
Of course, in the case $\gamma=0$ there is no $C$ state involved anymore and
the competition is only between the $A$ and $B$ states.
Moreover, as we note from Fig.~\ref{t_conv}, the convergence time to the one word consensus
state is the highest when $\gamma=0$ and decreases by increasing the
value of $\gamma$. This result is somewhat counter intuitive since
we expect that the presence of three states $A$, $B$, $C$ would slow
down the convergence with respect to a situation with only two states $A$ and $B$,
but actually in the first case another supplementary channel is yielding the stable $C$
case, i.e.\ the $AB\longrightarrow C$ channel (neglecting of course the rare $ABC\longrightarrow C$) thus accelerating the convergence.


The linearization of the mean field master equation around the absorbing points with $\beta>1/3$ delivers six negative eigenvalues, confirming that the points in the orange
and red region of Fig.~\ref{df-10000} are locally stable. Moreover, it comes out that
those eigenvalues do not depend on $\gamma$ showing that the choice of the initial conditions on $n_A$ and $n_B$ is crucial in entering the two different actraction basins. 
As a consequence of this independence on $\gamma$, the equation of the line dividing
the orange and red regions cannot be calculated easily.

\subsection{Low confidence \texorpdfstring{$\beta < 1/3$}{beta < 1/3}}

In the case $\beta<1/3$ we get multi-word final states. 
The green color in Fig.~\ref{df-10000} stands for an asymptotic situation where
$n_B=n_C$ and $n_{BC}=1-n_B-n_C$ (and of course a symmetric situation
with $B$ replaced by $A$ when the initial conditions favour $A$ rather than $B$).
The dependence of the asymptotic fractions $n_B$ and $n_C$ on $\beta$ 
is the same of that occurring for $\gamma=0$ and presented in \cite{BaronchelliBeta}.

\begin{figure}[!t]
\centering
\includegraphics[width=6cm]{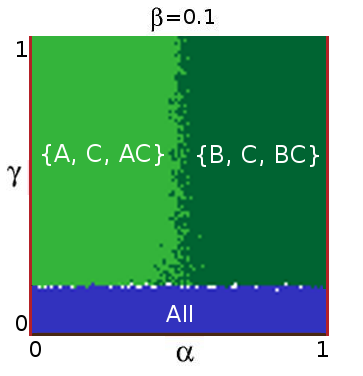}%
\includegraphics[width=6cm]{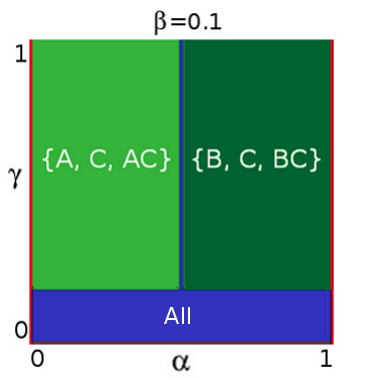}
\caption{
\textbf{\boldmath Phase diagram $\gamma,\alpha$ of the ABC model:}
Phase diagram depending on $\gamma$ and initial condition
$n_A=1-\alpha$ and $n_B=\alpha$ with $\alpha\in\left[ 0,1\right] $, with
$\beta=0.1$ fixed. 
\emph{Left:} diagram obtained after $t=10^6$ steps of the stochastic process. 
\emph{Right:} diagram obtained by integrating up to $t=1000$ the mean field 
master equation
with a fourth order Runge-Kutta algorithm employing a temporal step $dt=0.1$.
}
\label{g-al-b01}
\end{figure}

Instead, the blue color of Fig.~\ref{df-10000} represents an
asymptotic state where all vocabulary typologies are present $A$, $B$, $C$,
$AB$, $BC$, $AC$ and $ABC$, with $n_A=n_B$ and $n_{AC}=n_{BC}$. In this case
the vocabulary fractions depend both on $\beta$ and $\gamma$. 
White dots in the left panel of Fig.~\ref{df-10000}, which tend to disappear enlarging the population size $N$, are points
where the system has not shown a clear stable state after the chosen simulation
time. 
In fact, they disappear in the
final phase space described by the mean field master equation.
Contrary to the case of $\beta>1/3$, the final state does not depend on
the particular initial conditions provided that initially $n_A+n_B=1$. 
By fixing $\beta=0.1$ and varing $\gamma$ and initial conditions
$\alpha\in\left[0,1\right] $, we get the steady behavior shown in
Fig.~\ref{g-al-b01}.

The linearization of the mean field master equation around the absorbing points with $\beta<1/3$ in the green region of Fig.~\ref{df-10000} reveals that those are actractive points (six negative eigenvalues) irrespective of  the initial condition provided that 
$n_A+n_B=1$.
The equation of the transition line that divides the blue and green region can be
inferred numerically, again with the linearization of the master equation.
In particular the transition point at $\beta=0$ can be found analytically
to be at $\gamma=1/4$.
The independence on the initial conditions makes the region $\beta<1/3$ substantially
different from the complementary region $\beta>1/3$.

\section{Conclusions}

We modeled the emergence of a new language $C$ as a result of the mutual
interaction of two different populations of agents initially speaking different
idioms $A$ and $B$ and interacting each other without restriction (mean field).
Such tight connections between individuals speaking two different idioms
is certainly unrealistic, but the
same reasoning can be extended to accomplish the birth of single hybrid words resulting
from the interaction of two pronunciation variants of the same object (eg.\ the english word \emph{rowel}, perhaps an hybridization of the latin \emph{rota} and the english \emph{wheel}).

Three parameters govern the time evolution of the model and characterize the
final asymptotic state: 
$\beta$ the measure of the tendency of a hearer to adopt
the shared word used by the speaker (confidence), 
$\gamma$ the probability that two forms $A$
and $B$ are synthesized into the form $C$, 
and $\alpha$ the initial condition in the space $n_A+n_B=1$. 
It turns out that:
\begin{itemize}
\item for $\beta<1/3$
the system converges to multi-word states, all containing a fraction of the state
$C$, and that do not depend on the initial conditions provided that $n_A+n_B=1$.
\item for $\beta>1/3$
 the system converges to a consensus state where all agents end up
with the same state, either $A$, $B$ or $C$. 
The transition line $\gamma\left(\beta\right)$ separating the $A$
or $B$ convergence state from $C$, which are all locally stable independently
from $\gamma$, depends on the initial distribution of $n_A$ and $n_B$, with
$n_A+n_B=1$. Moreover, the invention of $C$ produces a reduction of
the time of convergence to the consensus state (all agents speaking with $C$)
when starting with an equal fraction of $A$ and $B$ in the population.
\end{itemize}
Interestingly the modern point of view of linguists links the birth and
continuous development of all languages as product of local interaction between
the varied language pools of individuals who continuously give rise to processes
of competition and exchange between different forms, but also creation of new
forms in order to get an arrangement with the other speakers \cite{Mufwene2001}.
In this view of a language as mainly a social product it seems that the use of
the Naming Game is particularly fit, in spite of the old conception of pure
languages as product of an innate psychological background of individuals \cite{Steels2011}.

It would be interesting to apply our model in the study of real language
phenomena where a sort of hybridization of two or more languages in a contact
language ecology takes ground. There are many examples of this in the history,
as for example the formation of the modern romance European languages from the
contact of local Celtic populations with the colonial language of Romans,
Latin. 
A more recent example of this is the emergence of Creole languages in colonial
regions where European colonialists and African slaves came into contact
\cite{Mufwene2007}. 

The starting point for a comparison of our model with this kind of phenomena
would be retrieving demographic data of the different ethnic groups at the moment they
joined in the same territory and observing if a new language established.
Our point of view would be obviously not to understand how particular speaking
forms emerged, but to understand whether there is a correlation between the
success of the new language forms and the initial language demography. 
In this case, a more refined modeling would take into account also the temporal
evolution of the population due to reproduction and displacements, and the
particular topologies related to the effective interaction scheme 
acting in the population. 

\section*{Acknowledgements}
The authors wish to thank V.~Loreto and X.~Castell\'o for useful discussions.
The present work is partly supported by the EveryAware european project 
grant nr.~265432 under FP7-ICT-2009-C.

\section*{Appendix: mean field master equation}

\label{appendix}
The mean field master equation of the ABC model, in the case
in which the speaker changes her vocabulary with the rule $\{AB, ABC\}
\stackrel{\gamma}{\longrightarrow}\{C\}$ before the interaction, is the
following:
{\tiny
\begin{eqnarray}
\frac{dn_A}{dt}&=&-n_An_B-n_Cn_A+
\left( \left( 1-\gamma\right) \frac{\beta-1}{2}+\beta-\gamma\right) 
 n_{AB}n_A+\frac{3\beta-1}{2}n_{AC}n_A-n_{BC}n_A+\nonumber \\
&&+\left(1-\gamma\right)\beta n_{AB}^{2}+\beta 
 n_{AC}^{2}+\beta\left( \left( 1-\gamma\right) 
 +1\right) n_{AB}n_{AC}+\left( \left( 1-\gamma\right)
 \frac{\beta-2}{3}+\beta-\gamma\right) n_An_{ABC}+\nonumber \\
&&+\beta\left( 1+\frac{2}{3}\left( 
 1-\gamma\right) \right) n_{ABC}n_{AC}+\left( 1-\gamma\right) \beta\left( 
 1+\frac{2}{3}\right)n_{ABC}n_{AB}+\left( 1-\gamma\right) \frac{2}{3}\beta
n_{ABC}^{2}
\nonumber \\
\frac{dn_B}{dt}&=&-n_An_B-n_Cn_B+\left( \left( 1-\gamma\right)
\frac{\beta-1}{2}+\beta-\gamma\right) 
 n_{AB}n_B+\frac{3\beta-1}{2}n_{BC}n_B-n_{AC}n_B+\nonumber \\
&&+\left(1-\gamma\right)\beta n_{AB}^{2}+\beta 
 n_{BC}^{2}+\beta\left( \left( 1-\gamma\right) 
 +1\right) n_{AB}n_{BC}+\left( \left( 1-\gamma\right) 
 \frac{\beta-2}{3}+\beta-\gamma\right) n_Bn_{ABC}+\nonumber \\
&&+\beta\left( 1+\frac{2}{3}\left( 
 1-\gamma\right) \right) n_{ABC}n_{BC}+\left( 1-\gamma\right) \beta\left( 
 1+\frac{2}{3}\right)n_{ABC}n_{AB}+\left( 1-\gamma\right) \frac{2}{3}\beta
n_{ABC}^{2}
\nonumber \\
\frac{dn_C}{dt}&=&-n_An_C-n_Bn_C+\frac{3\beta-1}{2}\left( n_{BC}+n_{AC}\right)
n_C-\left( 1-\gamma\right) n_{AB}n_C+\nonumber \\
&&+\gamma n_{AB}\left( 2\beta \left( n_{BC}+n_{AC}+n_{ABC}\right) +\left(
1-\beta \right) \left( n_{ABC}+n_{BC}+n_{AC}\right) +n_A+n_B+n_C+n_{AB}\right)
+\nonumber \\
&&+\beta n_{AC}^{2}+\beta n_{BC}^{2}+2\beta n_{AC}n_{BC}
+\left(\beta+\gamma+\left( 1-\gamma\right) \frac{\beta}{3}-\left(
1-\gamma\right) \frac{2}{3}\right) n_{ABC}n_C+\nonumber \\
&&+\gamma n_{ABC}\left( n_A+n_B+n_{AB}\right) +\left(2\beta\gamma+\left(
1-\gamma\right) \frac{2}{3}\beta+\left( 1-\beta\right) \gamma+\beta\right)
n_{ABC}\left( n_{AC}+n_{BC}\right)+\nonumber \\
&&+\left( \gamma\left( 1-\beta\right)+2\gamma\beta+\left( 1-\gamma\right)
\frac{2}{3} \beta\right) n_{ABC}^{2}
\nonumber \\
\frac{dn_{AB}}{dt}&=&2n_An_B+\frac{1}{2}n_{AC}n_B+\frac{1}{2}n_{BC}n_A-\left(
\gamma+1\right) n_{AB}n_C-\left(\left(  1-\gamma\right)
\frac{\beta-1}{2}+\beta+\gamma\right) n_{AB}\left( n_A+n_B\right)+\nonumber \\
&&-2\left( \beta\left( 1-\gamma\right) +\gamma\right) n_{AB}^{2}-\left(
\frac{1+\beta}{2}+\gamma+\left( 1-\gamma\right) \frac{\beta}{2}\right)
n_{AB}\left( n_{AC}+n_{BC}\right) +\nonumber \\
&&+\frac{1-\gamma}{3}n_{ABC}\left( n_A+n_B\right) -\left( \left( 1-\gamma\right)
\frac{2}{3}\beta+2\gamma+\frac{1-\gamma}{3}+\left( 1-\gamma\right) \beta\right)
n_{ABC}n_{AB}
\nonumber \\
\frac{dn_{AC}}{dt}&=&2n_An_C-2\beta n_{AC}^{2}-\frac{3\beta-1}{2}n_{AC}\left(
n_A+n_C\right) + \frac{1-\gamma}{2}n_{AB}n_C+\frac{1}{2}n_{BC}n_A+\gamma
n_{AB}n_A-n_Bn_{AC}+\nonumber \\
&&-\left( \beta\left( \gamma+\frac{1-\gamma}{2}+\frac{1}{2}\right)
+\frac{1-\gamma}{2}\right) n_{AB}n_{AC}-\frac{2\beta+1}{2}n_{AC}n_{BC}+\left(
\gamma+\frac{1-\gamma}{3}\right) n_{ABC}n_A+\nonumber \\
&&+\frac{1-\gamma}{3}n_{ABC}n_C-\left( \beta\left( \gamma+1+\frac{2}{3}\left(
1-\gamma\right) \right) +\frac{1-\gamma}{3}\right) n_{AC}n_{ABC}
\nonumber \\
\frac{dn_{BC}}{dt}&=&2n_Bn_C-2\beta n_{BC}^{2}\frac{3\beta-1}{2}n_{BC}\left(
n_B+n_C\right) + \frac{1-\gamma}{2}n_{AB}n_C+\frac{1}{2}n_{AC}n_B+\gamma
n_{AB}n_B-n_An_{BC}+\nonumber \\
&&-\left( \beta\left( \gamma+\frac{1-\gamma}{2}+\frac{1}{2}\right)
+\frac{1-\gamma}{2}\right) n_{AB}n_{BC}-\frac{2\beta+1}{2}n_{BC}n_{AC}+\left(
\gamma+\frac{1-\gamma}{3}\right) n_{ABC}n_B+\nonumber \\
&&+\frac{1-\gamma}{3}n_{ABC}n_C-\left( \beta\left( \gamma+1+\frac{2}{3}\left(
1-\gamma\right) \right) +\frac{1-\gamma}{3}\right) n_{BC}n_{ABC}
\nonumber \\
\frac{dn_{ABC}}{dt}&=&n_An_{BC}+n_Bn_{AC}+n_Cn_{AB}+\frac{1}{2}\left(
2n_{BC}n_{AC}+\left( 1+\left( 1-\gamma\right) \right) n_{AB}\left(
n_{AC}+n_{BC}\right) \right)+\nonumber \\
&&+\frac{1-\gamma}{3}n_{ABC}\left(n_{AB}+n_{AC}+n_{BC}\right)+\gamma
n_{AB}^{2}-\beta\left( n_A+n_B+n_C\right) n_{ABC}+\nonumber \\
&&-\beta\left( \left( 1-\gamma\right) n_{AB}+n_{AC}+n_{BC}\right)
n_{ABC}-\frac{1-\gamma}{3}\beta n_{ABC}\left( n_A+n_B+n_C\right)+\nonumber \\
&&-\frac{2\left( 1-\gamma\right) }{3}\beta
n_{ABC}\left(n_{AB}+n_{AC}+n_{BC}\right) -\gamma\beta n_{AB}n_{ABC}-\gamma
n_{ABC}\left( n_A+n_B+n_C\right)+\nonumber \\
&&-\gamma n_{ABC}\left( n_{AC}+n_{BC}\right)  -2\left( 1-\gamma\right) \beta
n_{ABC}^{2}-2\gamma\beta n_{ABC}^{2}-\gamma\left( 1-\beta\right) n_{ABC}^{2}
\label{eq:MF}
\end{eqnarray}
}

\end{document}